\documentclass{emulateapj}

\begin{document}

\slugcomment{ApJ accepted}

\title{The impact of galaxy geometry and mass evolution on the survival of star clusters}


\author{Juan P. Madrid, Jarrod R. Hurley, Marie Martig}

\affil{Center for Astrophysics and Supercomputing, Swinburne
University of Technology, Hawthorn, VIC 3122, Australia}

\begin{abstract}

Direct $N$-body simulations of globular clusters in a realistic Milky Way-like 
potential are carried out using the code \texttt{NBODY6} to determine the impact 
of the host galaxy disk mass and geometry on the survival of star clusters. 
A relationship between disk mass and star cluster dissolution timescale is 
derived. These $N$-body models show that doubling the mass of the disk from $5\times 10^{10}$~M$_{\odot}$ 
to $10\times10^{10}$~M$_{\odot}$ halves the dissolution time of a satellite star cluster 
orbiting the host galaxy at 6 kpc from the galactic center.
Different geometries in a disk of identical mass can determine either 
the survival or dissolution of a star cluster orbiting within the inner 
6 kpc of the galactic center. Furthermore, disk geometry has measurable effects 
on the mass loss  of star clusters up to 15 kpc from the galactic center. 
$N$-body simulations performed with a fine output time step show that at each 
disk crossing the outer layers of star clusters experience an increase in 
velocity dispersion of $\sim$5\% of the average velocity dispersion
in the outer section of star clusters. This leads to an enhancement of mass-loss -- 
a clearly discernable effect of disk shocking.
By running models with different inclinations we determine that star 
clusters with an orbit perpendicular to the Galactic plane have larger mass 
loss rates than both clusters evolving in the Galactic plane or in an inclined orbit.

\end{abstract}

\keywords{Galaxy: star clusters: general: galaxies: star clusters - galaxies: dwarf - 
galaxies-stars: evolution}

\maketitle

\section{Introduction}

In the current paradigm of galaxy formation smaller structures 
merge into larger ones from the Big Bang up to the present day 
(White \& Rees 1978).  Galaxies grow through two main processes: the 
hierarchical merging with smaller galaxies and the accretion of fresh 
gas fuelling new star formation. These different mechanisms contribute 
to the growth of the disk, bulge and halo.

As the first significant stellar structures to form, globular clusters witness 
the entire evolution of their host galaxy as satellite systems. Indeed, 
globular clusters are believed to  follow galaxies during galaxy 
mergers and close encounters (e.g.\ West et al.\ 2004). The gravitational 
potential of their host galaxy has a direct influence on the survival 
of globular clusters: they lose stars through tidal stripping and 
disk shocking. In turn, these lost stars contribute to the build-up of 
the galaxy's stellar halo. The evolution of host galaxy and globular 
clusters are clearly connected.

The aim of this work is to derive, using numerical simulations, the 
importance of the host galaxy disk mass and size on the evolution 
and survival of star clusters. How are satellite stellar systems affected 
by disks and bulges of changing mass and with different geometries? And 
reciprocally, how do satellite stellar systems contribute, through their 
dissolution, to the formation of the halo of the host galaxy?

D'Onghia et al.\ (2010) show that halo and disk shocking efficiently 
deplete the satellite population of dark matter halos within 30 kpc of
the Milky Way center. The results of that study cannot be directly applied 
to globular clusters because of their much higher densities compared to dark
matter halos.

In their important paper, Gnedin \& Ostriker (1997) model the dynamical
evolution of the Galactic globular cluster system using a Fokker-Planck 
code. These authors build on earlier analytical work by Aguilar, Hut \& 
Ostriker (1988) and Kundic \& Ostriker (1995) among others. Gnedin \& Ostriker (1997) 
give analytical expressions to estimate the impact of 
disk shocking and also call for numerical simulations to be carried out.  

With the recent progress in computational capacity, large $N$-body simulations
can now be carried out in order to determine the physical mechanisms that 
govern the dynamical evolution of globular clusters in a galactic potential.
Recently,  Renaud \& Gieles (2013)  found that after a merger event of 
the host galaxy the mass loss rate of a satellite star cluster increases. Interestingly, 
Renaud \& Gieles (2013) also find that even if the tidal forces reach a maximum during 
the merger itself they are too short-lived to have a significant impact on the long term survival 
of star clusters.

Previous work on the mass loss of star clusters focused on internal dynamical effects 
and stellar evolution. Vesperini \& Heggie (1997)
carried out numerical simulations of star clusters and determined their mass
loss rates through a Hubble time.
Baumgardt \& Makino (2003) established an analytical dissolution time scale for 
star clusters and showed that one third of the cluster mass is lost due to 
stellar evolution alone. A recent review by of recent $N$-body studies can be found in
Portegies Zwart et al.\ (2010).

The approach in our work is purely numerical and different in nature to 
Gnedin \& Ostriker (1997) given  that no assumptions or explicit expressions 
to treat the impact of the disk are used. 
$N$-body  models of star clusters were run where the properties of the star 
cluster remain identical but the mass and geometry of the galactic disk change. 
The effect of the disk on the star cluster is computed as a part of the numerical 
calculations of the gravitational force experienced by each star. 

The above is carried out with the code \texttt{NBODY6} used for the study of the 
dynamics of star clusters through $N$-body simulations. \texttt{NBODY6} now 
includes a detailed model of the host galaxy where star clusters evolve as 
satellites (Aarseth 2003). The current set-up of \texttt{NBODY6} includes the 
tools to model a Milky-Way type galaxy with three distinct components: disk, bulge, 
and halo. The gravitational force for each star of the cluster is computed 
at each time step by taking into account the effect of all other stars, and of 
the disk, bulge, and halo. We make use of this new capacity of \texttt{NBODY6} 
to run several models of star clusters where the mass and physical size of the 
disk are different between models. 

Throughout this work, a Hubble time of 13.5 Gyr is adopted, in agreement with
the results of the Wilkinson Microwave Anisotropy Probe (Spergel et al.\ 2003).


\section{Models set-up}

In the current framework of \texttt{NBODY6} the different components of the galactic 
model are static in time. Indeed the mass of the disk is assumed to be constant over time.  
In \texttt{NBODY6} the disk component of the galaxy is modeled following the 
prescriptions of Miyamoto and Nagai (1975):

\begin{equation}
\Phi(r,z)=\frac{GM_{DISK}}{\sqrt{r^2 +[a+\sqrt{(z^2+b^2)}]^2}}.
\label{eqn:eqmm}
\end{equation}

where $G$ is the gravitational constant and $M_{DISK}$ is the mass of the disk.
The geometry of the disk can be easily modified by changing the parameters
$a$ (scale length) and $b$ (scale height). Different values adopted in previous 
work for these two parameters and the disk mass are given in Table 1. 
While different values of these scale parameters are explored in this work, 
the most often assumed values are $a$~=~4~kpc and $b$ = 0.5 kpc. Also, a commonly 
used value for the mass of the bulge is $M_{BULGE}=1.5\times10^{10}M_{\odot}$  
(Xue et al.\ 2008) which we model as a point mass. We model the  galactic halo as a 
logarithmic potential that gives the entire galaxy a rotational velocity of 
220 km/s at 8.5 kpc from the galactic center (Aarseth 2003).


The initial set-up of the simulated star clusters is analogous to the simulations
presented in  Madrid et al.\ (2012). Briefly, the star clusters start with 
$N$=100~000 stars, an initial mass of $6.4\times10^{4}$ M$_{\odot}$, a half-mass 
radius of 6.2 pc, and a spatial distribution that assumes a Plummer 
sphere (Plummer 1911). The initial mass function (IMF) of the simulated star 
clusters is the one defined by Kroupa (2001), which extends the range given previously 
by Kroupa et al.\ (1993). This IMF defines 
a distribution of stellar masses using the quantity $\xi(m)dm$ which is the number of stars between 
the masses $m$ and $m + dm$. The explicit expression for $\xi(m)$ is the following
broken power law:

\begin{equation}
\xi(m) \propto m^{-\alpha_i},
\label{eqn:eqkkkk}
\end{equation}

where $\alpha_1 = 1.3$ for $0.08 \leq m/M_{\odot} < 0.5$ and  $\alpha_2 = 2.3$ for $0.5 \leq m/M_{\odot}$.
The lightest star in our simulation has a mass of 0.1 $M_{\odot}$ while the heaviest has a mass
of 50 $M_{\odot}$.
In addition, all simulations  have 5000 primordial binaries, that is 5\% of the total number of stars.

Models are run at 6 kpc from the galactic center, unless otherwise stated. At this 
distance, $R_{GC}$~=~6~kpc, star clusters are free from bulge shocking while the effect of disk 
shocking is still strong within the configuration described above. 
With a few clearly identified exceptions simulated star clusters follow 
a circular orbit with an initial inclination of $\theta =22.5$ degrees 
with respect to the galaxy disk. This  inclination gives the star cluster
an orbit that is neither planar nor perpendicular to the disk.
With this inclination the star cluster has a maximum height of $Z_{MAX}= 2$ kpc
comparable with the thick disk. In cartesian coordinates the plane of 
the disk is in the $(x,y)$ plane. 

Similarly to Madrid et al.\ (2012), stars are removed from the simulation when 
they have positive energy and when the distance from the center of the cluster
is at least two tidal radii. 
The tidal radius is defined as it was in Madrid et al.\ (2012), that is 
following the formula of K\"upper et al.\ (2010):\\

\begin{equation}
r_{\rm t} \simeq \left( \frac{GM_{\rm C}}{2\Omega^2} \right)^{1/3}
\label{eqn:eqkk}
\end{equation}

where $\Omega$ is the angular velocity of the cluster around the galaxy, 
$M_C$ is the mass of the cluster, and $G$ is the gravitational constant. 
We define an escape radius beyond which stars are no longer part of the 
simulation. This escape radius scales with the tidal radius over the time evolution 
of the cluster. The escape radius is at least two times the length of
the tidal radius.

The version of  \texttt{NBODY6} in use through this work is optimized to 
run on Graphic Processing Units (GPUs). All the models are performed on the GPU 
Supercomputer for Theoretical Astrophysical Research (gSTAR) hosted at 
Swinburne University. Each model is computed on one NVIDIA Tesla 
C2070 GPU in combination with six processing cores on the host node. 
The approximate computer time to carry out one model of $N$=100,000 stars 
is three weeks.\\


\begin{figure} 
\plotone{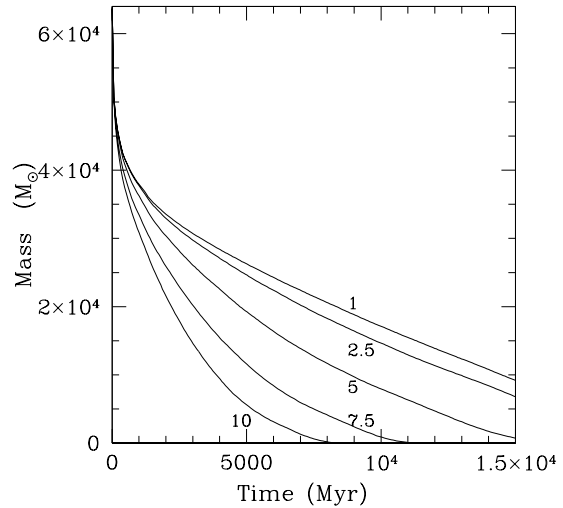} 
\caption{Total mass of simulated star clusters vs.\ time for models
with disks of different masses. The labels give the mass of the disk
in units of $10^{10}$ M$_{\odot}$, the lightest disk being $1\times 10^{10}$ M$_{\odot}$. 
Heavier disk masses enhance mass loss rates and accelerate the dissolution 
of star clusters. 
\label{totalmass}}
\end{figure}


\section{A heavier or lighter  disk}

Independent models of globular clusters are run with different masses for the
host galaxy disk. The mass of the disk is made to vary between $1\times10^{10}M_{\odot}$ 
and $10\times10^{10}M_{\odot}$ by incremental steps. The first step is of  $1.5\times10^{10}M_{\odot}$ 
while subsequent steps are of $2.5\times10^{10}M_{\odot}$. This range of values covers 
the different masses that a disk has during its evolution according to galaxy 
formation theory (e.g.\ Leitner 2012). These disk masses are also consistent with 
values published in the literature and shown in Table 1. This series of models 
are run at the same galactocentric distance of 6 kpc, and with the same 
properties, the only parameter that changes in the simulations is the disk mass.
The geometry of the disk is kept constant with $a=4$ kpc and $b=0.5$ kpc.

The total mass of simulated star clusters vs.\ time in simulations with different disk masses 
is plotted in Figure~\ref{totalmass}. A natural result of these simulations is that a more 
massive disk enhances the mass loss rate of an orbiting star cluster owing to a 
stronger tidal field. An enhanced mass loss rate implies a shortened dissolution time. 
The star cluster that orbits a ``light" disk with a mass of  $1\times10^{10}M_{\odot}$
has a remaining mass of $1.2\times10^{4} M_{\odot}$, or 19\% of its initial mass, 
after a Hubble time of evolution. On the other extreme, a star cluster that evolves 
within a ``heavy" disk with a mass of  $10\times10^{10}M_{\odot}$ is completely 
dissolved after 8.2 Gyr of evolution. As expected, disks with masses between these
two examples define intermediate regimes of mass loss, as shown in Figure~\ref{totalmass}.

\begin{figure} 
\plotone{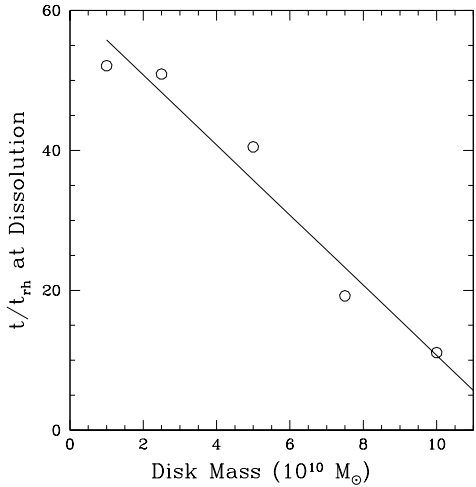} 
\caption{Dissolution time/relaxation time (t/t$_{rh}$) vs. disk mass for star 
clusters orbiting at 6 kpc from the galactic center. t/t$_{rh}$ is computed 
when the star cluster has only 10\% of its initial mass left. A heavier disk
shortens the number of relaxations a star cluster undergoes before dissolution.
\label{dissolutiontime}}
\end{figure}



\begin{center}
\begin{deluxetable}{lccc}
\tablecaption{Published Values for Mass and Structural Parameters of the Galactic Disk \label{tbl-1}} 
\tablewidth{0pt} 
\tablehead{
\colhead{Reference} & \colhead{Disk Mass (M$_{\odot}$)} & \colhead{$a$ (kpc)} & 
\colhead{$b$ (kpc)}
}
 
\startdata

Bullock \& Johnston (2005)   & $1.0\times 10^{11}$M$_{\odot}$  &  6.5  & 0.26 \\
G\'omez et al.\ (2010)      & $7.5\times 10^{10}$M$_{\odot}$  &  5.4  & 0.30 \\
Paczynski (1990)            & $8.1\times 10^{10}$M$_{\odot}$  &  3.7  & 0.20 \\
Pe\~narrubia et al.\ (2010) & $7.5\times 10^{10}$M$_{\odot}$  &  3.5  & 0.3  \\
Read et al.\ (2006)         & $5.0\times 10^{10}$M$_{\odot}$  &  4.0  & 0.50 \\

 \enddata
 
\tablecomments{This table gives published values for the mass and structural 
parameters of the disk for Milky-Way type galaxies. The values of $a$ and
 $b$ correspond to the disk scale length and disk scale height.}

\end{deluxetable}
\end{center}




By introducing the half-mass relaxation time $t_{rh}$, also computed by 
\texttt{NBODY6}, we can derive a simple relation between the number 
of relaxations a star cluster undergoes before dissolution 
and the mass of the disk. The half-mass relaxation time is given by 

\begin{equation}
t_{rh}=\frac{0.14N}{\ln\Lambda} \sqrt \frac{r_{hm}^3}{GM}, 
\label{eqn:eqnzz}
\end{equation}

where $\Lambda = 0.4N$ is the argument of the Coulomb logarithm, $N$ is the number 
of stars, and $r_{hm}$ the half-mass radius, (Spitzer \& Hart 1971; Binney \& 
Tremaine 1987). Figure~\ref{dissolutiontime} shows the number of relaxation times (t/t$_{rh}$)
vs. mass of the disk at the time when the cluster has only 10\% of its initial mass left.
This quantity ($10\% M_0$) is preferred over the time it takes for the cluster to completely dissolve 
since  numerical simulations become noisy for low $N$ owing to small number statistics. The best 
fitting linear relation between dissolution time, relaxation time and disk mass is plotted in Figure~\ref{dissolutiontime}
and is given by 

\begin{equation}
t/t_{rh}=-5\times \frac{M_{DISK}}{10^{10} M_{\odot}} + 61.
\label{eqn:eqzzz}
\end{equation}

We find that doubling the mass of the disk from $5\times10^{10}M_{\odot}$ to 
$10\times10^{10}M_{\odot}$ leads to the dissolution of
the star cluster in half the time. 


\section{Disk Geometry}

The current set-up of \texttt{NBODY6}, based on the formula of Miyamoto 
and Nagai (1975), allows the geometry of the disk to be modified by changing the 
values of the parameters $a$ and $b$ of Eq.\ 1. Table 1 shows that 
the mass and structural parameters of the Galactic disk take a 
range of values in different studies. Eighteen models of star clusters
where the  disk has different scale parameters were carried out in order to evaluate
the impact of disk geometry on the survival and evolution of star clusters. 
Selected parameters of models executed for this section are listed 
in Table 2. 

Three disk models are considered where the concentration of the disk mass
density profile varies from a disk mass highly concentrated towards the 
center of the galaxy to a disk mass with an extended mass profile.
Values for the disk scale parameters are $a=0.4$ kpc and $b=0.5$ kpc for 
the first set of models (models 1 through 6 in Table 2). 
By running models with $a=0.4$ kpc and $b=0.5$ kpc the shape of the disk 
changes to a more centrally concentrated one similar to a prolate bulge,
as the value of the scale length parameter $a$ is ten times smaller than 
the standard value used in the second set of models. The second set of models 
(labels 10 and 13 to 17 in Table 2) have disk parameters $a=4.0$ kpc and $b=0.5$ kpc. 
A third set of models was carried out with a ``flattened" or more extended disk 
where the scale parameters are $a=8$ kpc and $b=0.5$ kpc (models 18 to 23 in Table 2). 
Altogether, the evolution of 1.9 million stars was simulated for this section.

The disk density profiles of the three different disk geometries 
are represented in Figure 3.  The disk model with the Miyamoto 
and Nagai scale parameters $a=0.4$ kpc and $b=0.5$ kpc is represented on the 
top panel. In this model, the mass of the disk is highly 
concentrated towards the center of the galaxy with a very steep fall off:
the mid-plane disk density drops from 3.5$\times 10^9$ M$_{\odot}$/kpc$^3$ 
at R$_{GC}$ = 1 kpc to $1\times10^7$ M$_{\odot}$/kpc$^3$ at R$_{GC}$ = 7 kpc.

\begin{figure*} 
\plotone{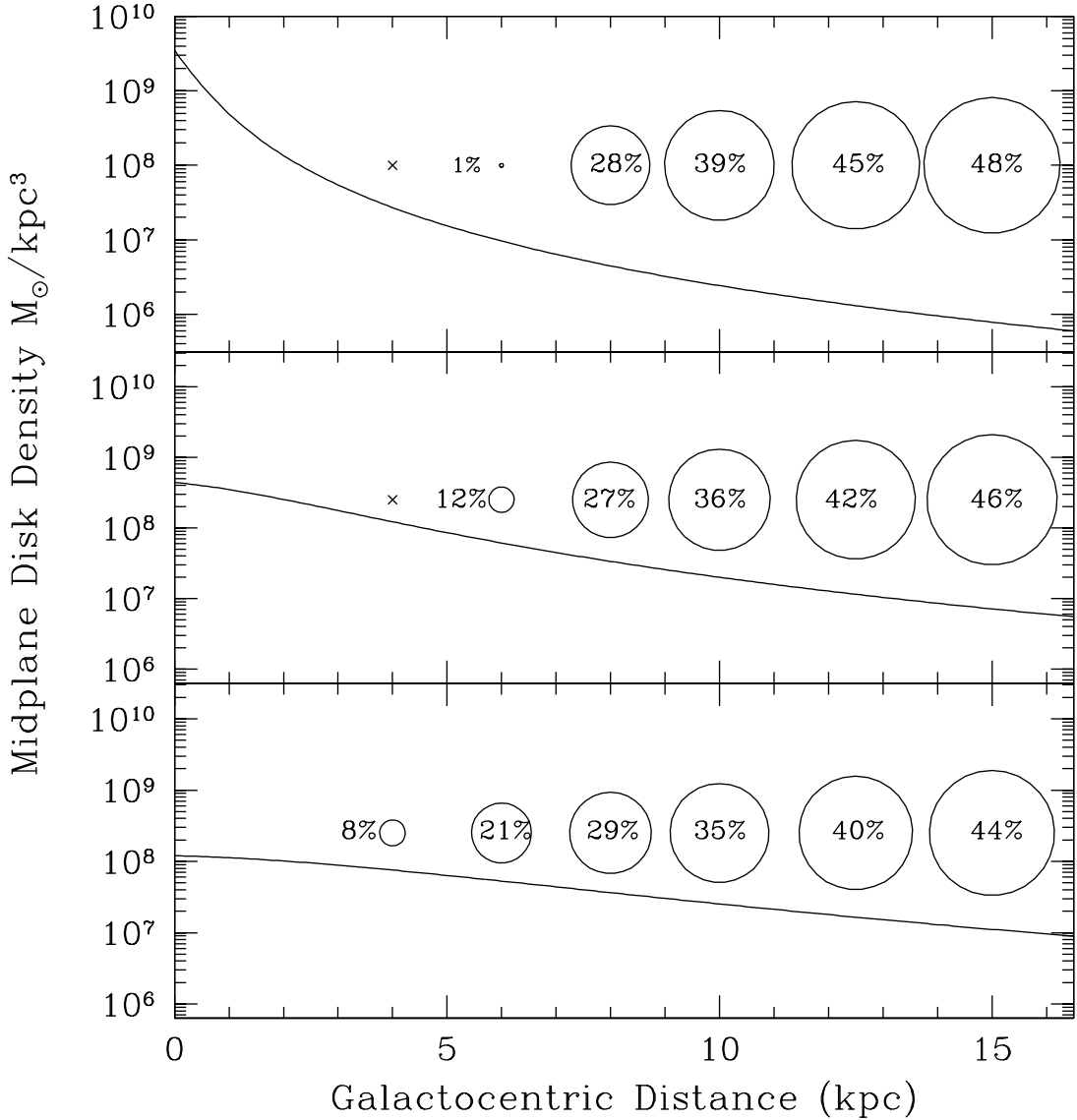} 
\caption{Effect of a different disk geometry on orbiting star clusters.
The mass density profiles, at the midplane ($z=0$), for three different disk 
geometries are plotted as a solid line in each panel. The size of the circles 
symbolizes the mass remaining in a star cluster after 10 Gyr of evolution. The percentage of the initial mass
left is also given for all models. Star clusters are represented at 
their respective galactocentric distances. {\it Upper panel}: A galaxy disk
with its mass concentrated at the center of the galaxy 
with scale length parameters of $a$=0.4 kpc and $b$=0.5 kpc. {\it Middle panel}: 
Galaxy disk with commonly used parameters $a$=4 kpc and $b$=0.5 kpc. {\it Lower panel}:
 A more extended galaxy disk, $a$=8 kpc and $b$=0.5 kpc.
\label{killerplot}}
\end{figure*}

For each disk profile we study the evolution of an identical star cluster 
at six different radii from the galactic center. Models were executed at 
R$_{GC}$ = 4, 6, 8, 10, 12.5, and 15 kpc from the galactic center. In 
Figure 3, circles with sizes proportional to the masses of the 
simulated star clusters after 10 Gyr of evolution are drawn at their 
respective orbital distances. 
The percentage of the initial star cluster mass remaining after 10 Gyr is also 
given in Figure~\ref{killerplot}.


Several simulated clusters do not survive to 10 Gyr, let alone up to a Hubble time.
For the first set of models, with a very concentrated disk, the simulated star cluster 
evolving at 4 kpc from the center of the galaxy is totally dissolved at a record time of only 
334 Myr. The simulated star cluster at an orbit of 6 kpc has a 
remaining mass of only $\sim$870 M$_{\odot}$ at 10 Gyr, just $\sim$1\% 
of its initial mass. At 8 kpc from the galactic center the modelled star 
cluster has 28\% of its initial mass after 10 Gyr of evolution.

The middle panel of Figure~\ref{killerplot} shows the disk density profile for
a disk model with $a=4$ kpc and $b=0.5$ kpc, i.e.\ the same geometry 
as the models of Section 3. In this second set of models, 
the star cluster orbiting at 4 kpc is also completely dissolved
before 10 Gyr. 

The star cluster orbiting at 6 kpc in this more extended disk model (middle panel) 
has a mass of 7972.8 M$_{\odot}$ after 10 Gyr of evolution, 
nine times the mass of the star cluster retained in a centrally peaked disk model. 
Clusters on orbits of 8 kpc in both models have virtually the same mass, i.e.\ only 
1\% difference, after 10 Gyr of evolution. Similarly, for models at $R_{GC}=10$ kpc 
and beyond their mass difference is clearly measurable but on the order of 3\% for 
clusters at $R_{GC}=10$ and 12.5 kpc and 2\% for clusters at $R_{GC}=15$ kpc. 

A third set of models, with a more extended disk, was also studied.
These models are represented in the bottom panel of Figure~\ref{killerplot}.
With this disk geometry, that is $a=8$ kpc and $b=0.5$ kpc, the mass 
of the disk is more spread out and the additive tidal effects of 
bulge and disk are not strong enough to completely disrupt the 
star cluster evolving at $R_{GC}=4$ kpc. This innermost star cluster 
survives 10 Gyr of evolution with 8\% of its initial mass. 

The models presented above show that disk geometry has a clear impact 
on the globular cluster mass function. This impact is more evident in 
the inner regions of the galaxy where a centrally concentrated disk 
(upper panel of Figure 3) adds to the tidal effects of the bulge and
enhances the destruction rate of globular clusters within 6 kpc of the 
galactic center. At $R_{GC}=8$ kpc there is a switch or transition
in the tidal effects following the different disk geometries. 
Star clusters at $R_{GC}=10$ kpc and beyond have more mass on the 
first set of models with the centrally 
concentrated disk than on the second and third set of models with more
extended disks. The tidal effects of the disk are still clearly measurable
out to $R_{GC}=15$ kpc but are small in intensity, of the order of
3 to 5\% of the initial mass. 

During 10 Gyr of evolution most of the mass of star clusters within 
R$_{GC} = 6$ kpc will be transfered to the host galaxy. Mass transfer 
from globular clusters to the host galaxy in the form of stellar tidal 
tails  has been well documented observationally in Galactic globular 
clusters and modelled by K\"upper et al.\ (2010). For instance, Odenkirchen 
et al.\ (2001) described the presence of two tidal tails emerging from Palomar 
5 using SDSS data. After dissolution, the stars that formed the star cluster 
become part of the host galaxy stellar population.


\section{Heating and cooling during disk crossings}

Gnedin \& Ostriker (1997) describe the effect of disk shocking as 
a shock or impulse of extra energy given to  each star in the cluster
as it crosses the disk (see also Spitzer 1958; Gnedin \& Ostriker 1999). 
Stars close to the tidal boundary can be lost during disk 
crossing events given that this extra energy allows them to escape 
from the star cluster. This section presents a close-up of the mass-loss
and velocity dispersion of star clusters at each disk crossing. 

A simulation with output given at more frequent intervals (every 1 Myr) was carried 
out to sample in detail the effects of a single disk crossing on a star cluster. The 
simulated star cluster is placed at 6 kpc from the galactic center with disk scale
parameters of $a$= 4 kpc and $b$= 0.5 kpc, an initial orbital inclination of $\theta =22.5$ degrees and 
an orbital period of $\sim$150 Myr. The height of  the star cluster above the plane 
of the disk and the internal velocity dispersion are represented in Figure 4. 
The internal velocity dispersion shown in Figure 4 corresponds to the 
velocity dispersion of the outer 50\% of the mass of the cluster, this is the 
section of the star cluster where individual stars experience the greatest 
changes in energy during each disk crossing. The velocity dispersion depicted
in Figure 4 has been offset from an average level of $\sim$2 km/s. For comparison, 
the average internal velocity dispersion of the entire star cluster is $\sim$3 km/s
after 3 Gyr of evolution. The maximum height reached by a star cluster is 2 kpc 
from the plane of the disk.

A periodic impulse given to the velocity dispersion of the outer layers of 
the star cluster at each disk crossing is evident in Figure 4. The amplitude
of this impulse, measured from an average level before disk crossing, is of
0.11 km/s. This increase is 5.2\% in the average velocity dispersion of the 
stars that make up the outer 50\% of the star cluster mass. There is a time delay 
of $\sim$7 Myr between the star cluster crossing the equatorial plane of 
the disk (i.e.\ $z$=0) and the peak of the impulse in velocity dispersion. 
This time delay is of the same order of magnitude as the crossing time
that is in this case $t_{cross} \sim 2$ Myr.
Following its peak the velocity dispersion of the star cluster 
experiences a reduction of $\sim0.14$ km/s on average. Gnedin et al.\ (1999) mention a 
``refrigeration effect" or slight reduction in the energy dispersion of a
star cluster following a disk shock. Figure 4 shows that at every disk 
crossing an increase is followed systematically by a decrease in velocity dispersion. 

The same velocity dispersion discussed above and the mass of the cluster 
outside its tidal radius ($M_{out}$) are  plotted in Figure 5. 
The mass outside the tidal radius $M_{out}$ plotted in Figure 5 
corresponds to the mass between the tidal radius and the escape radius.
The escape radius is at least twice the length of the tidal radius, as defined
in section 2 and in Madrid et al. (2012).

Figure 5 shows that at each disk crossing ($z$=0) the mass outside the 
tidal radius reaches a minimum, reflecting the strength of the tidal field
felt by the star cluster.  Over the ten disk crossings represented in 
Figures 4 the star cluster loses $\sim$2600 M$_{\odot}$, that is
on average 260 M$_{\odot}$ per disk crossing. 
After the star cluster passes the disk its velocity dispersion decreases 
and the mass outside the tidal radius increases as this region of the 
cluster fills up due to a weaker tidal field. $M_{out}$ reaches a maximum 
when the star cluster is at the furthest distance from the disk.

The stars that make up the mass outside the tidal radius come
from within the tidal radius as shown in Figure 5: the mass inside 
the tidal radius decreases when the mass outside the tidal radius increases.
For the time lapsed during the disk crossings represented in Figures 4 and 5, 
the tidal radius can be considered to be constant. Over longer time scales 
the tidal radius is a dynamic quantity by virtue of Equation 3. 


An elementary estimate for the mass loss during a disk crossing can
be derived using the data presented above and previous theoretical 
work. The energy change for stars in a star cluster is $\Delta E \sim (\Delta V)^2 $  
(Gnedin \& Ostriker 1997) and using the virial theorem $E~\propto~M^2$ we write that  
$dM/M \sim dV/V$.  
Using the simulations above we can give a numerical value to the 
expression $dM/M \sim dV/V$. The velocity increase in the outer regions 
of the star cluster has a numerical value of $dV/V \sim 0.1/2.0 = 0.05$ (see 
Fig.\ 5, top panel.) The change of mass can also be 
determined from the simulations, at 3 Gyr,  $dM/M=299/26125 \sim 0.01$. 
We find thus $dM/M \sim 0.2 dV/V$, that is, the change of mass is proportional
to roughly 20\% of the velocity dispersion change induced by a disk crossing.\\

\begin{figure} 
\plotone{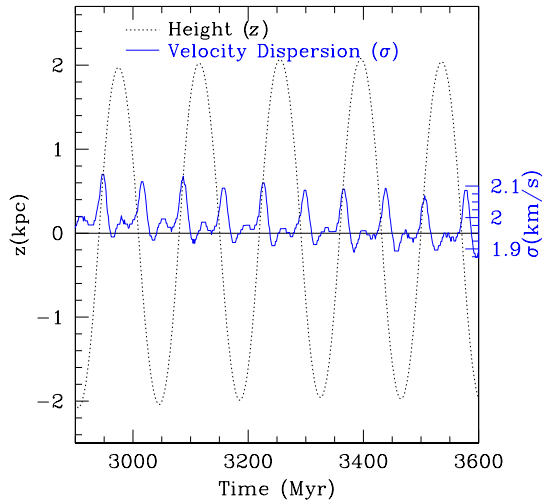} 
\caption{Star cluster height above the plane of the disk and velocity dispersion 
of stars in the outer 50\% of the star cluster mass.
At each disk crossing the outer layers of the star cluster experience an 
increase in their internal velocity dispersion. Both height and velocity 
dispersion have been scaled for display purposes. The average velocity dispersion 
decreases slightly over the time plotted in Figure 4. The maximum height in the orbit of the star 
cluster is 2 kpc above the plane of the disk.
\label{fig1}}
\end{figure}

\begin{figure} 
\plotone{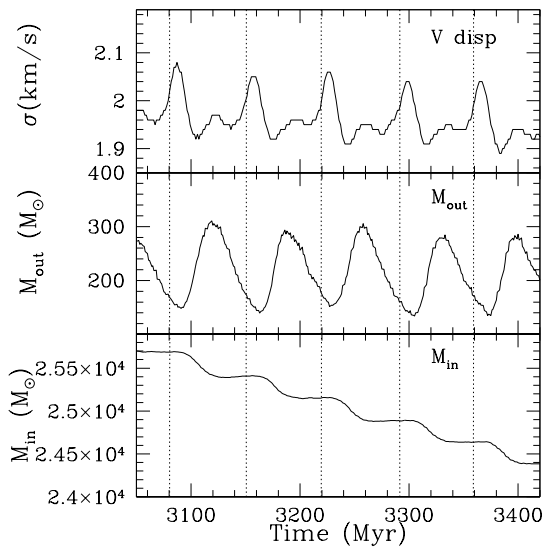} 
\caption{{\it Upper panel}: Velocity dispersion of stars in the outer 50\% of the
star cluster. {\it Medium panel}: Mass of the cluster outside the tidal radius ($M_{out}$). 
{\it Lower panel}: Mass inside the tidal radius ($M_{in}$). The vertical dotted lines
mark the crossing of the plane of the disk ($z=0$). During these five disk crossing
the tidal radius can be considered to be constant. The stars that make up the mass outside 
the tidal radius originate from within the tidal radius.
\label{fig1}}
\end{figure}

\begin{figure} 
\plotone{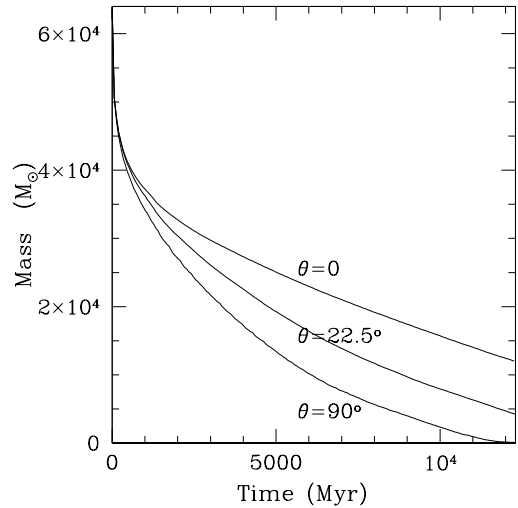} 
\caption{Star cluster mass vs. time for three clusters with different orbital
inclination. Three inclinations are shown here: a star cluster orbiting in the
plane of the disk $\theta = 0$, with a ``standard" orbit  $\theta = 22.5\degr$, and
perpendicular to the disk $\theta = 90\degr$.
\label{fig1}}
\end{figure}

\section{Orbital inclination}

For a cluster on an inclined orbit the passage through the disk implies an
enhanced, or impulsive, gravitational force owing to the proximity of the mass 
that constitutes the disk, as shown in the previous section. Is the transient
nature of the energy shift that stars receive during disk passages more 
important than a high but constant tidal field? In order to properly disentangle the 
tidal effects and those of disk shocking, two simulations with different orbital inclination
were carried out. One simulated star cluster orbits the galaxy in the plane 
of the disk, that is, all its initial velocity is in the $y$ direction 
$V=V_y, (\theta=0) $. A second simulation, with the same characteristics of the previous 
simulation, evolves in an orbit perpendicular to the disk ($\theta=90\degr$), 
in this case, $V=V_z$. Both simulations evolve at a galactocentric distance of 6 kpc, with 
a disk mass of $5\times10^{10} M_{\odot}$ and are otherwise identical to the
previous simulations with $\theta=22.5\degr$.

The impact of orbital inclination on the mass of the star cluster is shown
in Figure 6. We plot the total mass vs. time taking as a reference a simulated star 
cluster in a ``regular" or ``standard" orbit with an inclination of $\theta \sim  22.5^{\circ}$. 
We also plot the total masses of the simulated star clusters  with orbits on the plane of 
the disk ($\theta=0$), and perpendicular to it ($\theta=90\degr$).

Figure 6 shows that the star cluster in a perpendicular orbit to the 
plane of the disk is affected by a stronger mass loss than the cluster
in a less inclined orbit or the cluster in an orbit in the plane of the disk.
The simulated cluster on the orbit perpendicular to the disk is fully
dissolved in 12.3 Gyr. After 12 Gyr of evolution, the simulated cluster
evolving in the plane of the disk is 87 times more massive than the cluster
on the orbit perpendicular to the disk. That is a total mass difference 
of 4477 M$_{\odot}$.

The models discussed in this section show that the compressive shocks experienced
by the star cluster at each disk crossing are a dominant factor driving its 
dissolution. The variable gravitational potential generated by the presence of 
the disk on the orbit of the star cluster plays a more important role in the dissolution
of the cluster than a constant and higher tidal field experienced by a star cluster
that evolves in the plane of the disk. Note that star clusters with a small inclination 
angle with respect to the disk will experience additional shocking due to the presence of 
spiral arms (Gieles et al.\ 2007).


\begin{figure} 
\plotone{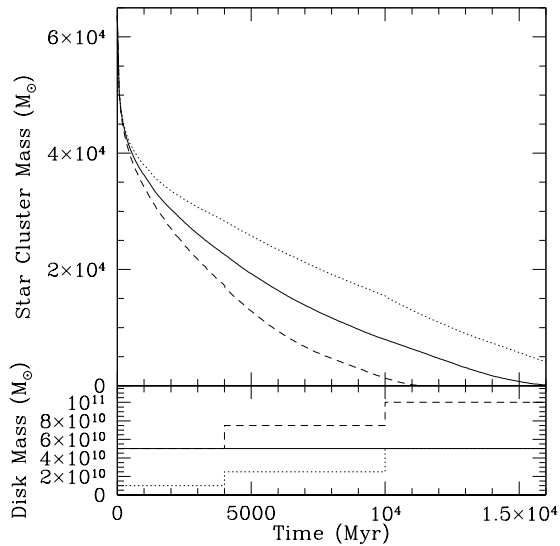} 
\caption{Star cluster mass vs.\ time for different evolution scenarios. The
accretion events experienced by the host galaxy can determine the survival 
or dissolution of a satellite star cluster. The solid line represents the 
star cluster evolving on a host galaxy with a constant disk mass of $5\times10^{10}$M$_{\odot}$.
The lower panel displays the evolution of the disk mass that increases over 
time with discrete accretion events.
\label{fig1}}
\end{figure}

\begin{center}
\begin{deluxetable*}{cccccccc}
\tablecaption{Parameters of Models Executed\label{tbl-1}} 
\tablewidth{0pt} 
\tablehead{
\colhead{Label} & \colhead{Bulge Mass} & \colhead{Disk Mass} & \colhead{$a$}
&  \colhead{$b$} & \colhead{$R_{GC}$} & \colhead{$t_{rh}$}  & \colhead{$t_{10\%}$}\\
\colhead{} & \colhead{(M$_{\odot}$)} & \colhead{(M$_{\odot}$)} & \colhead{(kpc)}
&  \colhead{(kpc)} & \colhead{(kpc)} & \colhead{(Gyr)}  & \colhead{(Gyr)}
}
\startdata
1    & $1.5\times10^{10}$ & $5\times10^{10}$    & 0.4   & 0.5   &  4    & 0.1 & 0.3 \\
2    & $1.5\times10^{10}$ & $5\times10^{10}$    & 0.4   & 0.5   &  6    & 1.1 & 7.2 \\
3    & $1.5\times10^{10}$ & $5\times10^{10}$    & 0.4   & 0.5   &  8    & 2.6 & -- \\
4    & $1.5\times10^{10}$ & $5\times10^{10}$    & 0.4   & 0.5   & 10    & 4.5 & -- \\
5    & $1.5\times10^{10}$ & $5\times10^{10}$    & 0.4   & 0.5   & 12.5  & 6.6 & -- \\
6    & $1.5\times10^{10}$ & $5\times10^{10}$    & 0.4   & 0.5   & 15    & 8.6 & -- \\
7    & $6.5\times10^{10}$ & 0                   & --    & --    &  6    & 8.7  & 6.1 \\
\hline

8    & $1.5\times10^{10}$ & $1.0\times10^{10}$  & 4.0   & 0.5   &  6    & 2.6 & 16.8 \\
9    & $1.5\times10^{10}$ & $2.5\times10^{10}$  & 4.0   & 0.5   &  6    & 2.3 & 15.3 \\
10    & $1.5\times10^{10}$ & $5.0\times10^{10}$  & 4.0   & 0.5   &  6    & 1.7 & 11.0 \\
11   & $1.5\times10^{10}$ & $7.5\times10^{10}$  & 4.0   & 0.5   &  6    & 1.2 & 6.8  \\
12   & $1.5\times10^{10}$ & $10\times10^{10}$   & 4.0   & 0.5   &  6    & 0.9 & 4.8  \\
13   & $1.5\times10^{10}$ & $5\times10^{10}$    & 4.0   & 0.5   &  4    &  0.9 & 5.1  \\
14   & $1.5\times10^{10}$ & $5\times10^{10}$    & 4.0   & 0.5   &  8    & 2.8 &  --    \\
15   & $1.5\times10^{10}$ & $5\times10^{10}$    & 4.0   & 0.5   & 10    & 4.0 &  --   \\
16   & $1.5\times10^{10}$ & $5\times10^{10}$    & 4.0   & 0.5   & 12.5  & 5.6 &  --   \\
17   & $1.5\times10^{10}$ & $5\times10^{10}$    & 4.0   & 0.5   & 15    & 7.2 &  --   \\
\hline
\hline
18    & $1.5\times10^{10}$ & $5\times10^{10}$    & 8.0   & 0.5   &  4    & 1.4 & 9.2\\
19    & $1.5\times10^{10}$ & $5\times10^{10}$    & 8.0   & 0.5   &  6    & 2.3 & 14.5\\
20    & $1.5\times10^{10}$ & $5\times10^{10}$    & 8.0   & 0.5   &  8    & 3.1 & 18.6\\
21    & $1.5\times10^{10}$ & $5\times10^{10}$    & 8.0   & 0.5   & 10    & 3.9 & 22.8 \\
22    & $1.5\times10^{10}$ & $5\times10^{10}$    & 8.0   & 0.5   & 12.5  & 5.0 & -- \\
23    & $1.5\times10^{10}$ & $5\times10^{10}$    & 8.0   & 0.5   & 15    & 6.3 & -- \\
 \enddata
\tablecomments{NOTE. -- Column 1 gives the model label; Column 2 bulge mass in solar masses; Column 3
disk mass in solar masses; Column 4: Miyamoto disk scale length $a$; Column 5 Miyamoto scale height $b$; 
Column 6 Galactocentric distance; Column 7: half-mass relaxation time ($t_{rh}$); Column 8: time when the star cluster
has only 10\% of the initial mass left $t_{10\%}$ -- Columns 7 and 8 are proxies for the dissolution time.}
\end{deluxetable*}
\end{center}


\section{Galaxy Evolution and the Survival or Dissolution of a Star Cluster}

At the present time \texttt{NBODY6} does not allow for a dynamic
host galaxy model. However, the first steps to include a time dependent 
potential have already been taken by Renaud et al.\ (2011). 
Within the current framework of \texttt{NBODY6}, and in order to investigate
the fate of a star cluster whose host galaxy grows with time, 
different simulations of star clusters evolving with different disk masses 
are combined together. These are the simulations presented in Section 3.

Two different pathways for the evolution of a star cluster are built.
The first series of accretion events experienced by the host galaxy leads to the 
survival of the star cluster after a Hubble time. The second scenario, where mass 
accretion lead to a more massive disk than in the first case, brings the star 
cluster to a complete dissolution before a Hubble time.

In the first scenario a simulated star cluster evolves during 4 Gyr as the satellite
of a host galaxy with a disk mass of $1\times 10^{10} M_{\odot}$. At 4 Gyr the
host galaxy undergoes an instantaneous accretion event that brings the mass of
the disk to $2.5\times 10^{10} M_{\odot}$. At 10 Gyr  the host galaxy 
experiences a second accretion event that brings the host galaxy disk to a mass
of $5\times 10^{10} M_{\odot}$. In this scenario, after a Hubble time, the star 
cluster has more mass than a simulated star cluster that evolves for a Hubble
time in a galaxy with a constant disk mass of $5\times 10^{10} M_{\odot}$, as expected.

In the second scenario the star cluster begins its evolution in a galaxy with a 
disk mass of $5\times 10^{10} M_{\odot}$, at  4 Gyr the disk mass increases
to $7.5\times 10^{10} M_{\odot}$ and at 10 Gyr it increases again to become
$10\times 10^{10} M_{\odot}$. The simulated star cluster is fully dissolved 
before a Hubble time due to the enhanced mass loss rates induced by these 
series of accretion events that built a more massive disk. Note that during this
exercise the geometrical parameters of the disk are kept constant $a=4$ and $b=0.5$ kpc.

The evolution of the total mass of the star cluster as a function of time in 
the two scenarios of accretion undergone by the host galaxy described above
are given in Figure 7. In addition to these two accreting models the 
mass evolution of a star cluster evolving around a disk with a constant 
mass of $5\times 10^{10} M_{\odot}$ is also plotted. This exercise shows how 
different accretion histories can lead to different depletion rates of satellite star clusters.

The results of this section can be obtained by joining together the appropriate 
curves from Fig.\ 1. For instance, we reconstruct a galaxy evolving in mass above
by joining the mass loss rates of a galaxy with a constant disk mass of $1, 2.5,$ and  $5\times 10^{10} M_{\odot}$
plotted in Fig.\ 1.
The results of this section are also in agreement with Equation 5 on 
Section 3 that yields a relation between dissolution time, relaxation time,
and the disk mass of the host galaxy. This equation shows, for instance,  that 
increasing the disk mass from  $1\times 10^{10} M_{\odot}$  to  $10\times 10^{10} M_{\odot}$ 
accelerates the destruction time of a star cluster from 56~$t_{rh}$ to 
11 $t_{rh}$.

We should note that we do not claim that the exercise above represents 
a realistic mass growth history of the Milky Way. Discrete accretion events
simulates what is certainly a smoothly growing galaxy. An exciting 
future perspective is to have upcoming versions of \texttt{NBODY6}
with a fully incorporated time dependent host galaxy potential.

\section{Final remarks}

The simulations presented in this work show that star clusters experience 
first hand the merger and accretion history of their host galaxy. The mass of
the host galaxy disk plays an important and measurable role in the evolution 
of satellite star clusters, by affecting their mass loss rates and thus their
structural parameters. The $N$-body models of star clusters have shown that
 different masses and geometries of the host galaxy disk 
can lead to different substructure within the inner 15 kpc of
the galactic center. With each galaxy having a different mass growth 
history, there is still a lot of work to understand how these
different histories affect globular clusters.
The mass and geometry of the disk  affect directly the
depletion rates of satellite stellar systems in a similar manner as dark matter 
halos are affected (D'Onghia et al.\ 2010).

Assuming a constant disk mass over a Hubble time, as it is often done, can
lead to an overestimate of the dissolution rates of globular clusters and thus 
impact the derived globular cluster mass function.


\acknowledgments
We would like to thank the referee for a constructive report that helped 
to improve this paper.
This  research has made use  of the NASA Astrophysics Data System
Bibliographic services (ADS) and  Google. This work was performed 
on the gSTAR national facility at Swinburne University of Technology. 
gSTAR is funded by Swinburne and the Australian Government’s Education 
Investment Fund. Many thanks to Darren Croton, Chris Flynn, Anna Sippel 
(Swinburne) and Allan Duffy (Melbourne University) for asking the 
inquisitive questions that inspired this work. 




\end{document}